\documentclass[twocolumn,showpacs,preprintnumbers,amsmath,amssymb,superscriptaddress]{revtex4-1}
\usepackage{graphicx}
\graphicspath{ {figs/} }

\newcommand{\gsim}{\;\rlap{\lower 3.5 pt \hbox{$\mathchar \sim$}} \raise 1pt
 \hbox {$>$}\;}
\newcommand{\lsim}{\;\rlap{\lower 3.5 pt \hbox{$\mathchar \sim$}} \raise 1pt
 \hbox {$<$}\;}

\newcommand{\Li}{{\rm Li}}

\usepackage{color}

\allowdisplaybreaks[4]

\begin{document}

\title{
\boldmath
The $O(\alpha^2)$ Initial State QED Corrections to $e^+e^-$ Annihilation \\ to a Neutral Vector Boson Revisited}

\author{J. Bl\"umlein}
\affiliation{Deutsches Elektronen--Synchrotron, DESY, Platanenallee 6, D--15738 Zeuthen, Germany}

\author{A. De Freitas}
\affiliation{Deutsches Elektronen--Synchrotron, DESY, Platanenallee 6, D--15738 Zeuthen, Germany}

\author{C.G. Raab}
\affiliation{Institute of Algebra, Johannes Kepler University, Altenbergerstra\ss{}e 69, A--4040, Linz, Austria}

\author{K. Sch\"onwald}
\affiliation{Deutsches Elektronen--Synchrotron, DESY, Platanenallee 6, D--15738 Zeuthen, Germany}

\date{\today}

\begin{abstract}
\noindent 
We calculate the non-singlet, the pure singlet contribution, and their interference term, at $O(\alpha^2)$ due to 
electron-pair initial state radiation to $e^+ e^-$ annihilation into a neutral vector boson in a direct analytic computation 
without any approximation. The correction is represented in terms of iterated incomplete elliptic integrals. Performing the limit 
$s \gg m_e^2$ we find discrepancies with the earlier results of Ref.~\cite{Berends:1987ab} and confirm results obtained in 
Ref.~\cite{Blumlein:2011mi} where the effective method of massive operator matrix elements has been used, which works for all
but the power corrections in $m^2/s$. In this way, we also confirm the validity of the factorization of massive partons in the 
Drell-Yan process. We also add non-logarithmic terms at $O(\alpha^2)$ which have not been considered in \cite{Berends:1987ab}. 
The corrections are of central importance for precision analyzes in $e^+e^-$ annihilation into $\gamma^*/Z^*$ at high luminosity. 
\end{abstract}
 
\preprint{
DESY 18--226, DO--TH 18/30}
 
\maketitle

\noindent The initial state QED corrections (ISR) to $e^+e^-$ annihilation are of crucial importance for the experimental 
analyzes at 
LEP \cite{EWWG} and for planned projects like the ILC and CLIC \cite{ILC}, the FCC\_ee \cite{FCCEE}, muon colliders 
\cite{Delahaye:2019omf}, and notably also at $e^+e^-$ Higgs factories using the process $e^+e^- \rightarrow Z H_0$, as well as  
$e^+e^- \rightarrow t \bar{t}$. The initial state corrections have been carried out analytically to $O\big((\alpha L)^5\big)$ in 
the 
leading logarithmic series using the structure function method \cite{LO}. A small $z$-resummation has been performed in 
\cite{Blumlein:1996yz}. In  Ref.~\cite{Berends:1987ab} the $O(\alpha^2)$ corrections were calculated 
neglecting terms of $O\big(\tfrac{m^2}{s} \ln(\tfrac{m^2}{s})\big)$. Here $s$ is the cms energy squared and $m = m_e$. These 
corrections 
are used in analysis codes such as {\tt TOPAZ0} \cite{Montagna:1998kp} and {\tt ZFITTER} \cite{ZFITTER}. The initial state QED 
corrections can be written in terms of the following functions 
\begin{eqnarray} 
H\left(z,\alpha,\frac{s}{m^2}\right) &=& \delta(1-z) + \sum_{k=1}^\infty \left(\frac{\alpha}{4\pi}\right)^k 
C_k\left(z, \frac{s}{m^2}\right) \\ C_k\left(z, \frac{s}{m^2}\right) &=& \sum_{l=0}^k \ln^{k-l}\left(\frac{s}{m^2}\right) 
c_{k,l}(z), 
\end{eqnarray} 
which yield 
the respective differential cross sections by 
\begin{eqnarray} 
\frac{d \sigma_{e^+e^-}}{ds'} = \frac{1}{s} \sigma_{e^+e^-}(s') H\left(z,\alpha,\frac{s}{m^2}\right), 
\end{eqnarray} 
with $\sigma_{e^+e^-}(s')$ the scattering cross section without the ISR corrections, 
$\alpha \equiv \alpha(s)$ the fine structure constant and $z = s'/s$, where $s'$ is the invariant mass of the produced 
(off-shell) $\gamma/Z$ boson.

In Ref.~\cite{Blumlein:2011mi} the method of massive operator matrix elements (OMEs) \cite{Buza:1995ie} has been applied to 
calculate the $O(\alpha^2)$ corrections, factorizing the process into universal massive contributions and 
the massless Wilson coefficients of the Drell--Yan process \cite{Hamberg:1990np,Harlander:2002wh}. This method has been 
known to work in the case of external massless fields, 
including the non-logarithmic contributions in QCD, cf.~\cite{Buza:1995ie,Blumlein:2016xcy}. However,
the results of Refs.~\cite{Berends:1987ab} and \cite{Blumlein:2011mi} were found to disagree.

One way to find out the correct answer is to perform the direct analytic calculation of the corresponding contributions without 
doing 
any approximation. This is done in the present paper for three of the subprocesses of the $O(\alpha^2)$ corrections related to 
fermion-pair production, i.e. for the processes II--IV of Ref.~\cite{Berends:1987ab}. Like in
Refs.~\cite{Berends:1987ab,Blumlein:2011mi} we will consider the case of pure vector couplings in the following. It is known 
already from the 
massless case \cite{Hamberg:1990np} that in some of the processes axialvector-vector terms receive different corrections if 
compared to the pure vector or axialvector case. These aspects will be presented in Ref.~\cite{BDRS}.
In the calculation we used the packages {\tt FORM, Sigma, HarmonicSums, 
HolonomicFunctions}
\cite{Vermaseren:2000nd,Schneider:2007a,Schneider:2013a,
Vermaseren:1998uu,Blumlein:1998if,
Ablinger:2014rba, Ablinger:2010kw, Ablinger:2013hcp, Ablinger:2011te, Ablinger:2013cf, 
Ablinger:2014bra,Ablinger:2017Mellin,KOUTSCHAN}
and private implementations \cite{RAAB1}.
The complete results have an iterative 
integral representation. We compare the exact result numerically with the one obtained in the limit $\rho = m^2/s \rightarrow 0$. 
Both results agree better than a relative deviation of $10^{-7}$ at $s = M_Z^2$, as expected by neglecting the power corrections. 
The result is given in terms of the variable $z$ and powers of the logarithm $L = \ln(s/m^2)$.  The logarithmic corrections in 
\cite{Berends:1987ab,Blumlein:2011mi} agree. 

In the calculation, the phase space integrals can be mapped to fourfold scalar 
integrals. Here one depends on $s''/s$, with $s''$ the mass squared of the emitted fermion pair. Unlike the case of 
$s$ and $s'$, the latter invariant is not large against $m^2$ everywhere. 
Three of the four integrals can be consecutively obtained yielding results which 
contain different functions whose arguments involve square-roots. It is then useful to construct a 
basis of the contributing root-valued
letters and to perform the last integral over it, using differential field methods \cite{Ablinger:2014bra}. Here also nested roots 
have to be transformed to single roots before.
By this the complete integrals are at most triple iterated over an
alphabet including also new types of square-root letters. Finally, the individual terms are regularized such that the expressions 
can be expanded in 
the ratio $\rho$ term by term \cite{BDRS1}. Typical letters are
\begin{eqnarray}
f_1(y; z, \rho) 
&=& \frac{y}{\sqrt{1-y} (w^2 + 4zy) w} \\
f_2(y; z, \rho) &=&
y/[(w^2+4zy)w\sqrt{y}\sqrt{-16\rho^2+(1-z)^2y}]
\nonumber\\
\end{eqnarray}
with
\begin{eqnarray}
&& w = \sqrt{16 \rho^2-8 \rho (1+z) y+(1-z)^2 y^2}.
\end{eqnarray}
Here $y$ denotes the next integration variable.
After having performed suitable regularizations, one may also expand in $\rho$ before performing the last integral.

Processes II and III have been also considered primarily for massless quark--antiquark pair production
in Refs.~\cite{Kniehl:1988id,SCHELLEKENS} in 4--dimensional calculations. Here a quark mass serves as a 
regulator since the massless limit is aimed at from the beginning. Neglecting the mass terms not needed to
regularize the corresponding integrals leads to simpler integrands, which finally integrate to
polylogarithms directly and the logarithmic contributions in $L$ are obtained correctly.
In the case discussed at present, however, the finite electron mass is physical and the expansion in $m^2$ 
is only possible if all terms which contribute to the final result are retained. This may require a deeper 
expansion in $m^2$ than the one performed in Ref.~\cite{Berends:1987ab}. Terms which can be safely neglected are
of $O\big(\rho \ln^k(\rho)\big),~k = 0,1,2$ in the result.

For process II in  \cite{Berends:1987ab} we find the difference term
\begin{eqnarray}
&& \frac{8}{3} \int_0^1 \frac{dy}{y} \sqrt{1-y} (2+y) \Biggl[\frac{(1-z)(1-(4-z)z)y}{4z+(1-z)^2 y}
\nonumber\\ && - \frac{1+z^2}{1-z} \ln \left(1 + \frac{(1-z)^2 y}{4z}\right)\Biggr] + O(\rho \ln(\rho)).
\end{eqnarray}
Here the original lower bound of the integral $4 m_e^2/(s (1-\sqrt{z})^2)$ can be set to zero since its contribution
is of the order of the neglected terms.
From this integral it follows that terms containing higher powers in $1/(1-z)$ are appearing, which were not contained
in \cite{Berends:1987ab,Kniehl:1988id}. They emerge from further terms in the mass expansion that cannot be neglected.

In the case of process III, Ref.~\cite{Berends:1987ab}, takes the results from \cite{SCHELLEKENS} in which, however, 
mass terms being necessary here, were neglected beforehand since the result concerned a massless quark calculation 
\cite{REM1}. Note also that the pure singlet interference term \cite{SCHELLEKENS} was taken with the wrong sign.

The difference terms, $\delta_i,~i =$ II, III, IV, for $e^+e^-$ pair emission between the present results and those of Ref. 
\cite{Berends:1987ab}, after the analytic expansion of the complete expression in $m^2/s$ including the constant term, read~: 
\begin{eqnarray}
\delta_{\rm II} &=& 
- \frac{128}{9} \Biggl[3 
+ \frac{1}{(1-z)^3} 
- \frac{2}{(1-z)^2} 
- 2 z\Biggr] 
-16 \Biggl[1 	
\nonumber\\ &&
+ \frac{5 z}{3} 
+ \frac{8}{9} \frac{1}{(1 - z)^4} - \frac{20}{9} \frac{1}{(1 - z)^3} 
+ \frac{4}{9} \frac{1}{(1 - z)^2} 
\Biggr] 
\nonumber\\ &&
\times \ln(z)
+ \frac{8}{3} \frac{1+z^2}{1-z} \Biggl[\frac{10}{9} - \frac{14}{3} \ln(z)  
-  \ln^2(z) \Biggr],
\\
\delta_{\rm III} &=& 
\frac{160}{3}-\frac{32}{z}+\frac{128}{3  (1+z)^2}
-\frac{64}{1+z}+96(1+z) \zeta_3
\nonumber\\ &&
-\Biggl[52(1-z)+\frac{64}{3z}(1-z^3)
\Biggr] \ln^2(z)
-\frac{56}{3}(1+z) 
\nonumber\\ &&
\times \ln^3(z)
+\Biggl[
24(1-z)
+16(1+z) \ln(z)\Biggr] \zeta_2
+ \ln(z) 
\nonumber\\ &&
\times\Biggl[
\frac{104}{3}
-\frac{32}{z}
+\frac{128}{3 (1+z)^3}
-\frac{256}{3 (1+z)^2}-\frac{64}{1+z}
\nonumber\\ &&
+64 \Biggl(1-z 
+ \frac{1-z^3}{3z}\Biggr) 
\ln(1+z)
\Biggr]
-\Biggl[40(1-z)
\nonumber\\ &&
+\frac{64}{3 z}(1-z^3)
+48(1+z) \ln(z)\Biggr] \Li_2(1-z)
\nonumber\\ &&
+64 \Biggl[1-z
+\frac{1}{3 z}(1-z^3)-(1+z) \ln(z)\Biggr] \Li_2(-z)
\nonumber\\ &&
+128(1+z) 
\Li_3(-z) -96(1+z) S_{1,2}(1-z)
\nonumber\\ &&
+ 2 \delta_{\rm interf}^{\rm PS},
\\
\delta_{\rm IV} &=&  
\frac{2 (53+994 z+32 z^2+742 z^3-85 z^4-8 z^5)}{9 (1-z) (1+z)^2}
\nonumber\\ &&
-8\Biggl[\frac{1-14 z-56 z^2+78 z^3-25 z^4}{(1-z^2)^2}
+\frac{1+z^2}{1-z}
\nonumber\\ &&
\times \ln(z)\Biggr] \zeta_2
-\frac{8z (13+12 z^2-20 z^3+3 z^4}{(1-z^2)^2} \ln^2(z)
\nonumber\\ &&
+16\Bigg[\frac{1-z+7 z^2-3 z^3}{(1+z)^2} 
+\frac{7+3 z^2}{2(1-z)}\ln(z)\Biggr] 
\nonumber\\ &&
\times \Li_2(1-z)
+ \Biggl[
\frac{32 (1+5 z-4 z^2)}{(1-z)^2} \ln(1+z)
\nonumber\\ &&
- \frac{16 (4 - 7  z- 6 z^2 - 128 z^3+2 z^4-9 z^5)}{3 (1-z)^2(1+z)^3} 
\Biggr] \ln(z)
\nonumber\\ &&
+\frac{32 (1+5 z-4 z^2)}{(1-z)^2} \Li_2(-z),
\end{eqnarray}
with $\delta_{\rm interf}^{\rm PS}$ the term given in (B.22) of \cite{Hamberg:1990np} setting $C_F = T_F = 1, \alpha_s = \alpha$.
Here, $\Li_k(z)$ and $S_{p,k}(z)$ denote the polylogarithm and Nielsen integrals, respectively, cf. \cite{Devoto:1983tc}.
We remark that we agree with the result  on the interference terms in the pure singlet case \cite{SCHELLEKENS} which 
has been found there already to be regularization scheme invariant; it also agrees with the massless result
\cite{Hamberg:1990np}. Numerically, it turns out that the deviation due to the $\delta_{\rm interf}^{\rm PS}$ is small against 
the other differences for the pure singlet term. We agree with the result for $\mu^+\mu^-$ emission for the 
non-singlet term (process II) of Refs.~\cite{Berends:1987ab,Kniehl:1988id}, see also Ref.~\cite{BDRS}. This contribution has also 
a 
representation using  massive OMEs. Here, however, the external fermion lines are massless since $m_e \ll m_{\mu}$, see 
Ref.~\cite{QCD3}. The terms of $O(1/(1+z)^{2(3)})$ given in \cite{Berends:1987ab} for process IV have not been observed in 
the respective contribution of the OMEs in \cite{Blumlein:2011mi}. Terms of this kind are only expected for 
(anti)particle-(anti)particle scattering. 

\begin{figure}[th]
  \centering
  \hskip-0.8cm
  \includegraphics[width=.9\linewidth]{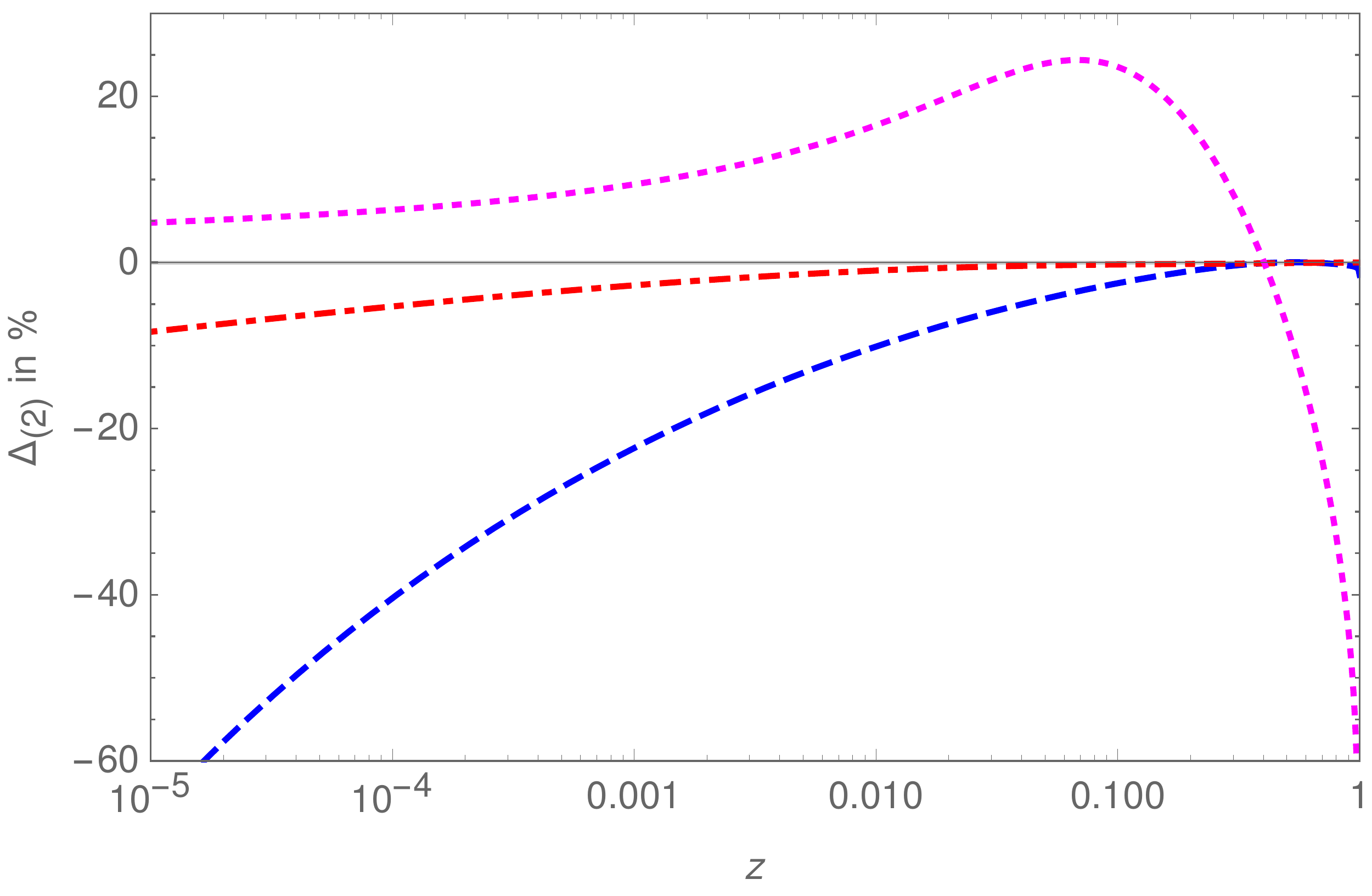}
  \caption{Relative deviations of the results of Ref.~\cite{Berends:1987ab} from the exact result in \% for the $O(\alpha^2)$ 
  corrections. 
  The non--singlet contribution (process II): dash-dotted line;
  the pure singlet contribution (process III): dashed; 
  the interference term between both contributions (process IV): dots; for $s = M_Z^2$, $M_Z = 91.1879$~GeV.} 
  \label{fig:reldev}
\end{figure}

The relative deviations for the results for processes II--IV in the present calculation and \cite{Berends:1987ab} are shown 
in Figure~\ref{fig:reldev}, where $\Delta_{(2)}$ denotes the ratio of $\delta_i$ and the corresponding complete
$O(\alpha^2)$ correction for $i =$  II, III, IV. All illustrations are made for $z < 1$. The relative differences reach from 
+25 to --60\% for $z \in [10^{-5},1]$. Here we have changed the term $\ln(z)/(1-z)^2 \rightarrow \ln^2(z)/(1-z)^2$ in 
Eq.~(2.43) in \cite{Berends:1987ab} which appears twice (suggesting a typo), such that this term is only logarithmic but not
linear divergent for $z \rightarrow 1$ and thus integrable.
Otherwise the difference would be even larger.

\begin{figure}[th]
  \centering
  \includegraphics[width=.9\linewidth]{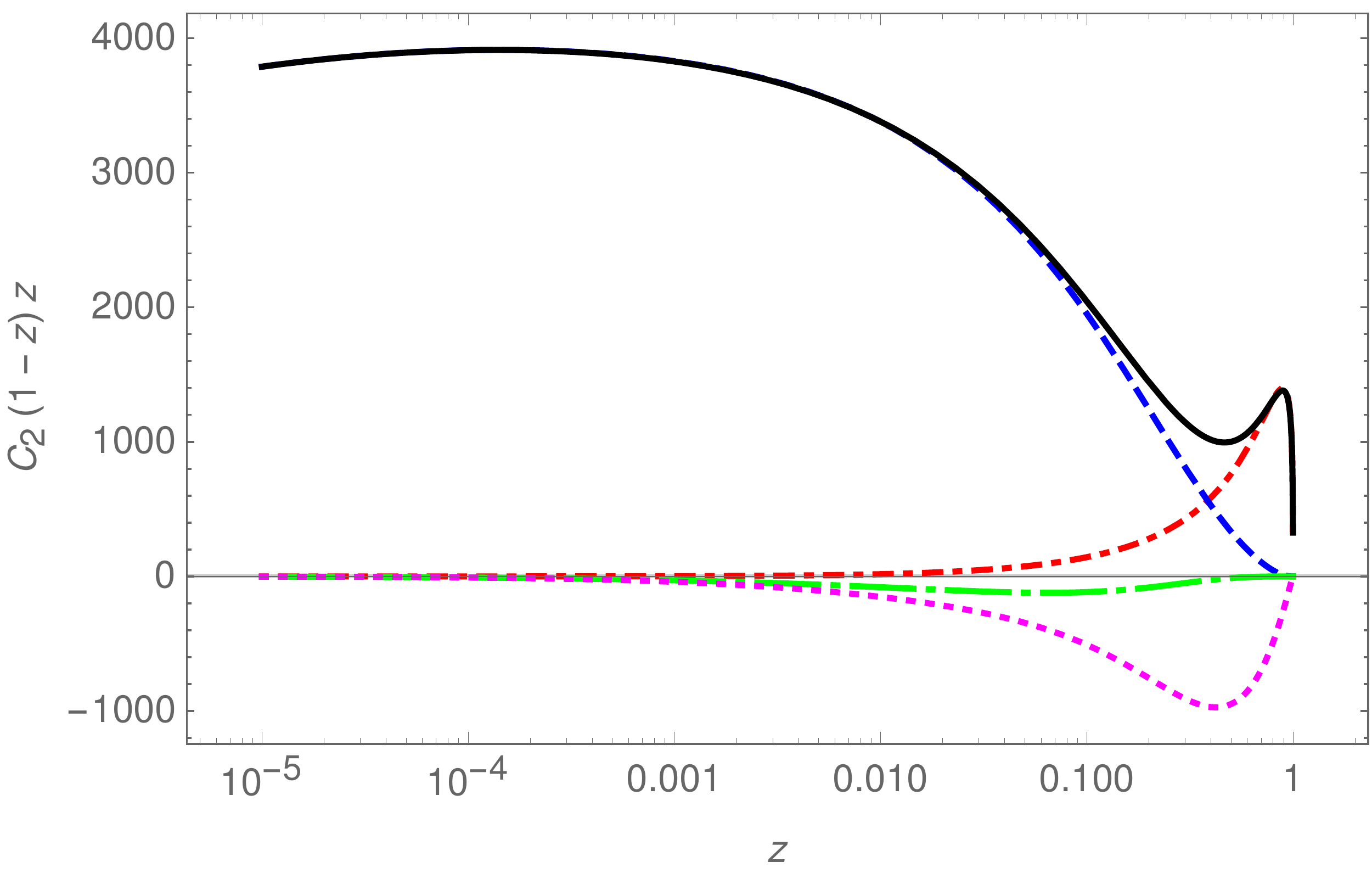}
  \caption{The initial state $O(\alpha^2)$ corrections to $\gamma^*/Z^*$ production due to $e^+ e^-$ pair production multiplied by 
$z(1-z)$. 
  The non--singlet contribution (process II): dash-dotted line;
  the pure singlet contribution (process III): dashes; 
  the interference term between both contributions 
  (process IV) $\times 10$: dotted; the vector contributions implied by process B, Ref.~\cite{Hamberg:1990np}, and interferences 
$\times 100$: 
long 
  dash-dotted; all contributions: full line for $s = M_Z^2$.} 
  \label{fig:fermcorr}
\end{figure}

For the non-logarithmic terms, not all contributions have been considered in \cite{Berends:1987ab}. 
These are the graphs B  and their interference terms with the non-singlet (A) and pure singlet terms (C and D) in
\cite{Hamberg:1990np}. We have recalculated them in the massive case. Note that the interference term between 
the graphs A and B only contributes to the axialvector term. For these processes there is no massive OME.
Since the massive OMEs contain all massive corrections in the limit $m^2/s \rightarrow 0$, the only contributions
are from the massless Wilson coefficients. We find in the corresponding massive calculation the massless results 
given in \cite{Hamberg:1990np}, which is a further confirmation of the formalism presented in Ref.~\cite{Buza:1995ie} 
including the constant terms. This has also been observed e.g.~in the case of the massive asymptotic two--loop corrections
to the deep--inelastic structure function $F_L(x,Q^2)$ \cite{Buza:1995ie,Blumlein:2006mh}.
Figure~\ref{fig:fermcorr} shows the different contributions at $O(\alpha^2)$ of initial state $e^+e^-$ pair 
production to $\gamma^*/Z^*$-boson production. The dominant contributions come from the pure singlet and non--singlet terms; 
other contributions are smaller but not negligible at the 0.1\% level in the radiator function. For large values of 
$z = s'/s$ the non--singlet terms are dominant, whereas for $z \lsim 0.03$ the pure singlet contributions dominate.

Analogous contributions to those considered here have been calculated at three--loop order for the 
massive OMEs in QCD with external massless parton lines in \cite{QCD3}. When Ref.~\cite{Blumlein:2011mi} was published, 
we could not explain the differences to Ref.~\cite{Berends:1987ab} and we tended to assume that the OME method might have 
a problem in case of massive external states, which is now proven not to be the case. The factorization of massive initial 
states for the Drell-Yan process \cite{Collins:1998rz} is also observed in the case discussed here. 
\section*{Acknowledgments}

\noindent 
This paper is dedicated to the memory of our colleague W.L.~van Neerven. We would like to thank J.C.~Collins, 
J.H.~K\"uhn, G.~Passarino, C.~Schneider and G.~Sterman for discussions. This project has received funding from the 
European Union’s Horizon 2020 research and innovation programme under the Marie Sk\/{l}odowska-Curie grant agreement No. 
764850, SAGEX, and COST action CA16201: Unraveling new physics at the LHC through the precision frontier 
and from the Austrian FWF grants P 27229 and P 31952
in part.


\begin{thebibliography}{99}
\bibitem{BDRS1}
The complete results are lengthy. They will be given, together with computational 
details, in \cite{BDRS}.
\end{thebibliography}
\end{document}